\documentclass[aip,jap,twocolumn,superscriptaddress,longbibliography,10pt]{revtex4-1}

\usepackage{glossaries}
\usepackage{graphicx}
\usepackage[usenames,dvipsnames,svgnames]{xcolor}
\usepackage{hyperref}
\hypersetup{
pdfnewwindow=true,          
colorlinks=true,            
linkcolor=DarkBlue,         
citecolor=DarkBlue,         
filecolor=DarkBlue,         
urlcolor=DarkBlue           
}

\newacronym{aa}{AA}{all-atom}
\newacronym{cg}{CG}{coarse-grained}
\newacronym{dft}{DFT}{density-functional theory}
\newacronym{md}{MD}{molecular dynamics}
\newacronym{mlp}{MLP}{machine-learned potential}
\newacronym{nep}{NEP}{neuroevolution potential}
\newacronym{pmf}{PMF}{potential of mean force}
\newacronym{rdf}{RDF}{radial distribution function}
\newacronym{rmse}{RMSE}{root-mean-square error}

\begin{document}

\title{NEP-CG and NEP-AACG: Efficient coarse-grained and multiscale all-atom-coarse-grained neuroevolution potentials}

\author{Zheyong Fan}
\email{brucenju@gmail.com} 
\affiliation{College of Physical Science and Technology, Bohai University, Jinzhou, P. R. China}
\affiliation{Suzhou Laboratory, Suzhou, Jiangsu 215123, P. R. China}

\author{Wenjun Zhang}
\affiliation{College of Physical Science and Technology, Bohai University, Jinzhou, P. R. China}

\author{Zhenhao Zhang}
\affiliation{Key Laboratory of Material Simulation Methods \& Software of Ministry of Education, College of Physics, Jilin University, Changchun 130012, China}

\author{Ke Xu}
\email{kickhsu@gmail.com}
\affiliation{College of Physical Science and Technology, Bohai University, Jinzhou, P. R. China}

\author{Xuecheng Shao}
\email{shaoxc@jlu.edu.cn}
\affiliation{Key Laboratory of Material Simulation Methods \& Software of Ministry of Education, College of Physics, Jilin University, Changchun 130012, China}

\author{Haikuan Dong}
\email{donghaikuan@163.com}
\affiliation{College of Physical Science and Technology, Bohai University, Jinzhou, P. R. China}

\date{\today}

\begin{abstract}
Machine-learned coarse-grained (CG) models often suffer from noisy training data, limiting their accuracy and transferability. We propose a method to generate low-noise training data based on the potential of mean force by constraining CG beads during atomistic simulations and accumulating time-averaged forces. Implemented within the neuroevolution potential (NEP) framework, our approach achieves training accuracy comparable to atomistic models trained on density functional theory data. For liquid water, the NEP-CG model accurately reproduces densities from 1 bar to 1 GPa, successfully extrapolating beyond the 0.5 GPa training limit, with a virial correction essential for the correct equation of state. For an anisotropic C$_{60}$ monolayer, distinguishing crystallographically distinct bead types reduces stress errors by an order of magnitude and captures directional thermal conductivity. We further introduce a multiscale NEP-AACG model integrating all-atom (AA) and CG degrees of freedom, demonstrated for gold nanowire fracture at an experimentally relevant strain rate. Computational speeds for NEP-CG models reach hundreds to thousands of ns/day using a single consumer-grade GPU. This work provides a robust framework for constructing accurate, transferable, and efficient CG models across diverse systems.
\end{abstract}

\maketitle

\section{Introduction}

\Gls{md} simulations have become indispensable in chemistry, biology, and materials science, providing atomistic insights into complex systems. However, a persistent challenge is the vast range of time and length scales governing molecular phenomena. Many processes occur on microsecond or longer timescales, far exceeding those accessible to conventional \gls{aa} simulations. \Gls{cg} modeling addresses this by grouping atoms into fewer interaction sites, or ``beads'', reducing the effective degrees of freedom and enabling simulations of significantly larger systems and longer timescales \cite{noid2013jcp}. The central goal of \gls{cg} models is to preserve the essential structural and thermodynamic properties of the original atomistic system.

The force-matching method, proposed by Izvekov and Voth \cite{Izvekov2005jpcb} and placed on a rigorous statistical mechanics foundation by Noid \textit{et al.} \cite{Noid2008jcp}, provides a systematic approach for optimizing \gls{cg} model parameters. Subsequent developments have produced various tools and \gls{md} engines for constructing and deploying \gls{cg} models \cite{peng2023jpcb, Mirzoev2019cpc, anderson2020hoomd, zhu2013jcc, xu2025mgea_pygamd}.

When a \gls{cg} model is derived by formally integrating out degrees of freedom, multi-body interactions naturally emerge in the effective energy function. Including such multi-body terms has been shown to improve model accuracy \cite{molinero2009jpcb,larini2010jcp,wang2021jcp}. The advent of \glspl{mlp} \cite{unke2021machine} offers new opportunities, as \glspl{mlp} excel at capturing many-body effects in complex systems.

Consequently, there has been growing interest in using \glspl{mlp} to learn \gls{cg} models from \gls{aa} \gls{md} trajectories. John and Csanyi \cite{john2017jpcb} developed a \gls{cg} model based on the Gaussian approximation potential framework \cite{bartok2010prl}, demonstrating significantly higher accuracy than pair-potential models while remaining faster than \gls{aa} simulations for solvent-free biomolecular systems. Zhang \textit{et al.} \cite{zhang2018jcp} constructed a DP-CG water model using the force-matching method and the deep potential approach \cite{zhang2018prl}, accurately reproducing oxygen structural characteristics. Wang \textit{et al.} \cite{wang2019acscs} introduced CGnets, showing that \glspl{mlp} can capture explicit-solvent free energy surfaces with only a few \gls{cg} beads, unlike classical methods. \gls{mlp}-based \gls{cg} methods have since been extended to proteins \cite{majewski2023nc, charron2025nc} and other complex systems. Beyond local invariant-feature-based \glspl{mlp}, approaches using graph neural networks \cite{husic2020JCP, ruza2020jcp, thaler2022jcp} and equivariant features \cite{loose2023jpcb} have also been explored for constructing \gls{cg} models.

Despite these advances, current MLP-based \gls{cg} methods often rely on training data with substantial noise. Reported \gls{rmse} values for typical water models \cite{loose2023jpcb} range from 3.49 to 11.8 kcal/mol/\AA~(0.151--0.512 eV/\AA), significantly higher than typical \gls{aa} \gls{mlp} errors for water (e.g., $\sim$0.05 eV/\AA) \cite{zhang2018prl}. This noise causes several problems: (1) large training errors obscure convergence assessment; (2) noise complicates hyperparameter selection; (3) many training structures are required even for a single state point; and (4) resulting models are often applicable only to a single density and lack transferability across different pressures.

To address these challenges, we propose a method for generating low-noise training data based on the potential of mean force definition. After thermal equilibration, we constrain the bead degrees of freedom during \gls{aa} \gls{md} simulations and accumulate time-averaged forces. These averages correspond to the true mean forces on the beads and are inherently smooth, making them easy for an \gls{mlp} to learn.

Our implementation uses the \gls{nep} framework \cite{fan2021prb} within the GPUMD package \cite{xu2025mgea}, one of the most computationally efficient \gls{mlp} architectures available. NEP has found widespread application in \gls{aa} simulations \cite{ying2025cpr} and recently in \gls{cg} contexts \cite{argun2025jcp, li2025npj}. By reducing training data noise, we achieve accuracy comparable to \gls{aa} models trained on \gls{dft} data, resolving the issues above: (1) high accuracy enables clear convergence assessment; (2) optimal hyperparameters can be selected based on training accuracy; (3) only one or a few structures suffice for a robust model at a single state point; and (4) accurate virial data enable description of various strained states, conferring transferability across strain conditions. Notably, we find that \gls{cg} models require larger cutoff radii than their \gls{aa} counterparts but do not need many trainable parameters.

Beyond pure \gls{cg} simulations, we introduce a multiscale NEP-AACG model that combines \gls{aa} and \gls{cg} degrees of freedom within a single framework. This integration is natural for NEP, which handles many-component systems efficiently \cite{song2024nc, liang2025arxiv}.

\section{Methods}
 \label{section:methods}
 
\subsection{Neuroevolution potentials}
\label{sec:nep}

Our \gls{cg} and AACG models are based on the \gls{nep} approach~\cite{fan2021prb}, specifically its fourth-generation version (NEP4)~\cite{song2024nc} implemented in the GPUMD package~\cite{xu2025mgea}. NEP4 is particularly suitable for multi-component systems.

\gls{nep} uses a feedforward neural network to represent the site energy $U_i$ of atom $i$ as a function of a descriptor vector $\mathbf{q}$ with $N_\mathrm{des}$ components:
\begin{equation}
U_i(\mathbf{q}) = U_i \left(\{q^i_{\nu}\}_{\nu=1}^{N_\mathrm{des}}\right).
\end{equation}
The network has a single hidden layer with $N_\mathrm{neu}$ neurons and $\tanh$ activation:
\begin{equation}
U_i = \sum_{\mu=1}^{N_\mathrm{neu}} w^{(1)}_{\mu} \tanh\left(\sum_{\nu=1}^{N_\mathrm{des}} w^{(0)}_{\mu\nu} q^i_{\nu} - b^{(0)}_{\mu}\right) - b^{(1)},
\end{equation}
where $\mathbf{w}^{(0)}$ and $\mathbf{w}^{(1)}$ are connection weights, and $\mathbf{b}^{(0)}$ and $b^{(1)}$ are biases.

The descriptor for atom $i$ comprises radial and angular components constructed from the local atomic environment within a cutoff distance. The radial components are:
\begin{equation}
q^i_{n} = \sum_{j\neq i} g_{n}(r_{ij}), \quad 0\leq n\leq n_\mathrm{max}^\mathrm{R},
\end{equation}
giving $n_\mathrm{max}^\mathrm{R}+1$ radial components. Angular components include up to five-body terms. For three-body terms:
\begin{equation}
q^i_{nl} = \sum_{m=-l}^l (-1)^m A^i_{nlm} A^i_{nl(-m)},
\end{equation}
with
\begin{equation}
A^i_{nlm} = \sum_{j\neq i} g_n(r_{ij}) Y_{lm}(\theta_{ij},\phi_{ij}),
\end{equation}
where $Y_{lm}$ are spherical harmonics, $0\leq n\leq n_\mathrm{max}^\mathrm{A}$, and $1\leq l \leq l_\mathrm{max}^\mathrm{3b}$. Four-body and five-body terms follow Ref.~\cite{fan2022gpumd}.

The radial functions $g_n(r_{ij})$ appear in both radial and angular descriptors, expanded as:
\begin{equation}
g_n(r_{ij}) = \sum_{k=0}^{N_\mathrm{bas}^\mathrm{R}} c^{ij}_{nk} f_k(r_{ij}),
\end{equation}
where
\begin{equation}
f_k(r_{ij}) = \frac{1}{2}\left[T_k\left(2\left(r_{ij}/r_\mathrm{c}^\mathrm{R}-1\right)^2-1\right)+1\right] f_\mathrm{c}(r_{ij}),
\end{equation}
with $T_k$ Chebyshev polynomials and $f_\mathrm{c}(r_{ij})$ a cosine cutoff:
\begin{equation}
f_\mathrm{c}(r_{ij}) =
\begin{cases}
\frac{1}{2}\left[1 + \cos\left( \pi \frac{r_{ij}}{r_\mathrm{c}^\mathrm{R}} \right) \right], & r_{ij}\leq r_\mathrm{c}^\mathrm{R}; \\
0, & r_{ij} > r_\mathrm{c}^\mathrm{R}.
\end{cases}
\end{equation}
For angular descriptors, $N_\mathrm{bas}^\mathrm{R}$ and $r_\mathrm{c}^\mathrm{R}$ are replaced by $N_\mathrm{bas}^\mathrm{A}$ and $r_\mathrm{c}^\mathrm{A}$.
Importantly, atom-type information is encoded directly in the expansion coefficients $c_{nk}^{ij}$, enabling efficient handling of multi-component systems without separate descriptor branches.

Parameters are optimized using the separable natural evolution strategy (SNES)~\cite{schaul2011high}. The loss function combines RMSE for energies, forces, and virials with $\mathcal{L}_1$ and $\mathcal{L}_2$ regularization, weighted by tunable hyperparameters $\lambda_{\rm e}$, $\lambda_{\rm f}$, $\lambda_{\rm v}$, $\lambda_1$, and $\lambda_2$. Energies and virials are in eV/atom, forces in eV/\AA.

\subsection{Force-matching for coarse-graining}

The force-matching method, originally developed for extracting classical effective forces from \textit{ab initio} simulations~\cite{ercolessi1994el}, was adapted for \gls{cg} models~\cite{Izvekov2005jpcb} and given a rigorous statistical foundation by Noid \textit{et al.}~\cite{Noid2008jcp}. The idea is to derive a \gls{cg} potential such that forces on \gls{cg} beads approximate the ensemble-averaged forces on corresponding atomistic groups.

Consider an atomistic system with $N_{\rm AA}$ atoms with coordinates $\mathbf{r}$ and potential energy $U(\mathbf{r})$. The system is mapped onto $N_\mathrm{CG}$ \gls{cg} beads defined by a linear mapping $\mathbf{R} = \mathbf{C} \cdot \mathbf{r}$. Each bead $I$ represents a group of atoms, with position typically given by the center of mass:
\begin{equation}
\mathbf{R}_I = \sum_{i \in I} C_{Ii} \mathbf{r}_i, \quad C_{Ii} = \frac{m_i}{\sum_{i' \in I} m_{i'}}.
\end{equation}
In the canonical ensemble, the equilibrium distribution of \gls{cg} coordinates is:
\begin{equation}
P(\mathbf{R}) = \left\langle \delta(\mathbf{R} - \mathbf{C} \cdot \mathbf{r}) \right\rangle \propto \int d\mathbf{r} \, \delta(\mathbf{R} - \mathbf{C} \cdot \mathbf{r}) e^{-\beta U(\mathbf{r})}.
\end{equation}
This distribution can be generated by a \gls{cg} model with potential $U_\mathrm{CG}(\mathbf{R})$ satisfying $P(\mathbf{R}) \propto e^{-\beta U_\mathrm{CG}(\mathbf{R})}$, identifying $U_\mathrm{CG}(\mathbf{R})$ as the \gls{pmf}:
\begin{equation}
\label{equation:pmf}
U_\mathrm{CG}(\mathbf{R}) = -k_\mathrm{B}T \ln \left[ \int d\mathbf{r} \, \delta(\mathbf{R} - \mathbf{C} \cdot \mathbf{r}) e^{-\beta U(\mathbf{r})} \right].
\end{equation}
The PMF depends on the thermodynamic state point, reflecting integrated-out degrees of freedom.

From the PMF, the mean force on bead $I$ at \gls{cg} configuration $\mathbf{R}$ is:
\begin{equation}
\label{equation:cg_force}
\mathbf{F}_I^\mathrm{CG}(\mathbf{R}) = -\nabla_{\mathbf{R}_I} U_\mathrm{CG}(\mathbf{R}) = \left\langle \sum_{i \in I} \mathbf{f}_i \right\rangle_{\mathbf{R}},
\end{equation}
where $\mathbf{f}_i$ are atomic forces and $\langle \cdots \rangle_{\mathbf{R}}$ denotes an ensemble average constrained by $\mathbf{C} \cdot \mathbf{r} = \mathbf{R}$. Thus, target forces for a \gls{cg} model are constrained ensemble averages of total forces on each bead's atoms.

Traditional force-matching minimizes the following loss function~\cite{Izvekov2005jpcb}:
\begin{equation}
\label{equation:chi2}
\chi^2 = \left\langle \frac{1}{3N_\mathrm{CG}} \sum_{I=1}^{N_\mathrm{CG}} \left\| \sum_{i \in I} \mathbf{f}_i(\mathbf{r}) - \mathbf{F}_I^\mathrm{CG}(\mathbf{C} \cdot \mathbf{r}) \right\|^2 \right\rangle,
\end{equation}
where $\mathbf{F}_I^\mathrm{CG}(\mathbf{R})$ are predicted \gls{cg} forces. A crucial distinction exists between Eq.~(\ref{equation:cg_force}) and Eq.~(\ref{equation:chi2}): the former defines the ensemble-averaged forces, while the latter optimizes the \gls{cg} model against instantaneous forces. This approximation in the traditional force-matching method works well for models with limited flexibility. However, flexible \glspl{mlp} could be prone to overfitting the substantial noise in instantaneous forces, especially with limited data, a challenge that is only partially mitigated by standard regularization techniques.

\subsection{NEP-CG and NEP-AACG methods}

\subsubsection{NEP-CG method}

Building on atomistic NEP and force-matching, we develop NEP-CG and NEP-AACG approaches. The key innovation is generating low-noise training data that directly correspond to the \gls{pmf} via constrained \gls{md} simulations, rather than fitting to instantaneous forces.

For each thermodynamic state, we first quilibrate the atomistic system at target temperature and pressure/density using atomistic NEP. 
Then we apply constraints to fix the \gls{cg} bead positions and continue the simulation in the NVE ensemble, accumulating instantaneous forces (Eq.~(\ref{equation:cg_force})) and virial on each bead $I$.
Time-averaged quantities converge to true mean values as $\sim 1/\sqrt{T}$ (with $T$ production time).
In practice, we found that a production time of 1--10 ns is sufficient to achieve a level of noise notably lower than the attainable accuracy for the \gls{nep} approach.

A notable feature in our approach is that we also use virial stresses as training targets, which is crucial for obtaining \gls{cg} models transferrable to different pressure states.
For a \gls{cg} bead $I$, the instantaneous virial is the sum of constituent atom virials and the ensemble average is:
\begin{equation}
\label{equation:virial}
\mathbf{W}_I = \left\langle \sum_{i\in I} \mathbf{W}_i\right\rangle_{\mathbf{R}},
\end{equation}
with the atomistic virial contribution from atom $i$ given by~\cite{fan2021prb}
\begin{equation}
\mathbf{W}_i = \sum_{j\neq i} \mathbf{r}_{ij} \otimes \frac{\partial U_j}{\partial \mathbf{r}_{ji}}.
\end{equation}
Here $\mathbf{r}_{ij}\equiv \mathbf{r}_{j}-\mathbf{r}_{i}$.
The total virial in the \gls{cg} system is 
\begin{equation}
    \mathbf{W} = \sum_I \mathbf{W}_I.
\end{equation}
We found that training directly with this virial definition systematically underpredicts pressure and overpredicts density, because coarse-graining eliminates kinetic (ideal gas) contributions from integrated-out atoms. 
To remedy this, we introduce a virial correction: 
\begin{equation}
\label{equation:virial_correction}
\mathbf{W} \to \mathbf{W} +  (N_{\mathrm{AA}} - N_{\mathrm{CG}}) k_{\mathrm{B}}T \mathbf{I},
\end{equation}
where $\mathbf{I}$ is the identity tensor. This correction compensates for lost degrees of freedom, ensuring correct pressure response.

For pure \gls{cg} simulations, NEP-CG models are trained to reproduce ensemble-averaged quantities from constrained simulations. The \gls{nep} architecture (Section~\ref{sec:nep}) is unchanged, but site energies $U_i$ represent free energy contributions of \gls{cg} beads. Training data include  bead forces (Eq.~\ref{equation:cg_force}) and virial tensor. 
We will show that minimizing the NEP loss function with these targets yields accuracy comparable to atomistic NEP models trained on \gls{dft} data.
The method is also data efficient.
For one thermodynamic state, only one or a few configurations suffice for a robust NEP-\gls{cg} model, unlike conventional approaches requiring thousands of snapshots.
Consequently, we still have a relative small dataset for training a \gls{cg} model applicable to a wide range of pressures states. 

\subsubsection{NEP-AACG method}

We further introduce NEP-AACG, which integrates atomistic and \gls{cg} degrees of freedom within a single model for multiscale simulations. The NEP architecture handles mixed-resolution systems by treating atomistic sites and \gls{cg} beads as distinct species. Descriptors are constructed from local environments that may include both types, with the same radial and angular framework applying uniformly. Different particle pairs may require different cutoff radii to optimally capture local environments.

Training data are generated by extending constrained dynamics: for each configuration, all sites (atomistic and \gls{cg}) are constrained, and ensemble-averaged quantities on every site are accumulated. The NEP-AACG model learns a consistent free energy surface spanning both resolutions by predicting all quantities simultaneously.

Coupling between atomistic and \gls{cg} regions emerges naturally from training data, without \textit{ad hoc} schemes. The unified NEP framework enables smooth transitions between regions and supports systems with spatial resolution variation. NEP-AACG thus provides a unified, data-driven approach to multiscale modeling.

\section{Results and discussion}
\label{section:results}

We demonstrate the capabilities of the NEP-CG and NEP-AACG methods through three representative examples. First, we use liquid water, one of the most extensively studied systems in both traditional \gls{cg} and MLP-CG approaches, to illustrate the key features of our NEP-CG method and introduce a virial correction scheme essential for accurate density prediction. 

Second, we consider a solid-state system: a monolayer of covalently bonded C$_{60}$ molecules in the quasi-hexagonal phase. Here we examine the effects of coarse-graining on vibrational properties and demonstrate how the NEP-CG model captures the essential dynamics while significantly reducing computational cost.

Third, we showcase the NEP-AACG method through a study of gold metal. We construct a mixed-resolution model and apply it to investigate the deformation of gold nanowires under tension, where the central region is modeled atomistically while surrounding regions are coarse-grained to extend accessible length and time scales.

\subsection{NEP-CG model for liquid water}

To construct NEP-CG models for liquid water, we select a NEP-AA model from our previous work \cite{xu2025npj}. This atomistic model was trained against CCSD(T)-level MB-pol reference data and accurately predicts structural, thermodynamic, and transport properties across a wide range of conditions.

\subsubsection{Training data generation}

We first perform atomistic \gls{md} simulations using the NEP-AA model in the NPT ensemble to achieve thermal equilibrium at 300 K and pressures from 1 bar to 0.5 GPa. Specifically, we select discrete pressure points: 1 bar, 10 bar, 0.1 GPa, 0.2 GPa, and 0.5 GPa. For each pressure, we conduct a 1 ns NPT simulation for equilibration, then switch off the thermostat and barostat while applying constraints to the \gls{cg} beads. Each \gls{cg} bead corresponds to the center of mass of the three atoms in a water molecule. We perform a 10 ns NVE simulation to accumulate force and virial data. The atomistic system contains 1536 atoms, mapping to 512 \gls{cg} beads. 
The timestep for integration is 0.5 fs.
The complete training dataset comprises just 5 structures (2560 water beads in total), confirming the high data efficiency of our approach.

\subsubsection{Model hyperparameters}

The potential energy surface of a \gls{cg} model is expected to be smoother than its atomistic counterpart due to integrated-out degrees of freedom. Consequently, the \gls{cg} model requires substantially fewer trainable parameters. In the NEP approach, the number of trainable parameters is primarily controlled by the number of hidden-layer neurons $N_{\rm neu}$. For the NEP-AA model \cite{xu2025npj}, $N_{\rm neu}=60$; through extensive testing, we found $N_{\rm neu}=10$ sufficient for the NEP-CG model, confirming the expected smoothness.

Cutoff radii for radial and angular descriptors influence both accuracy and performance. Since the water molecule center of mass is comparable in size to an oxygen atom, we adopt the same cutoffs as the NEP-AA model: radial cutoff 6~\AA, angular cutoff 4~\AA. 

\subsubsection{Model accuracy}

\begin{figure}
    \centering
    \includegraphics[width=\linewidth]{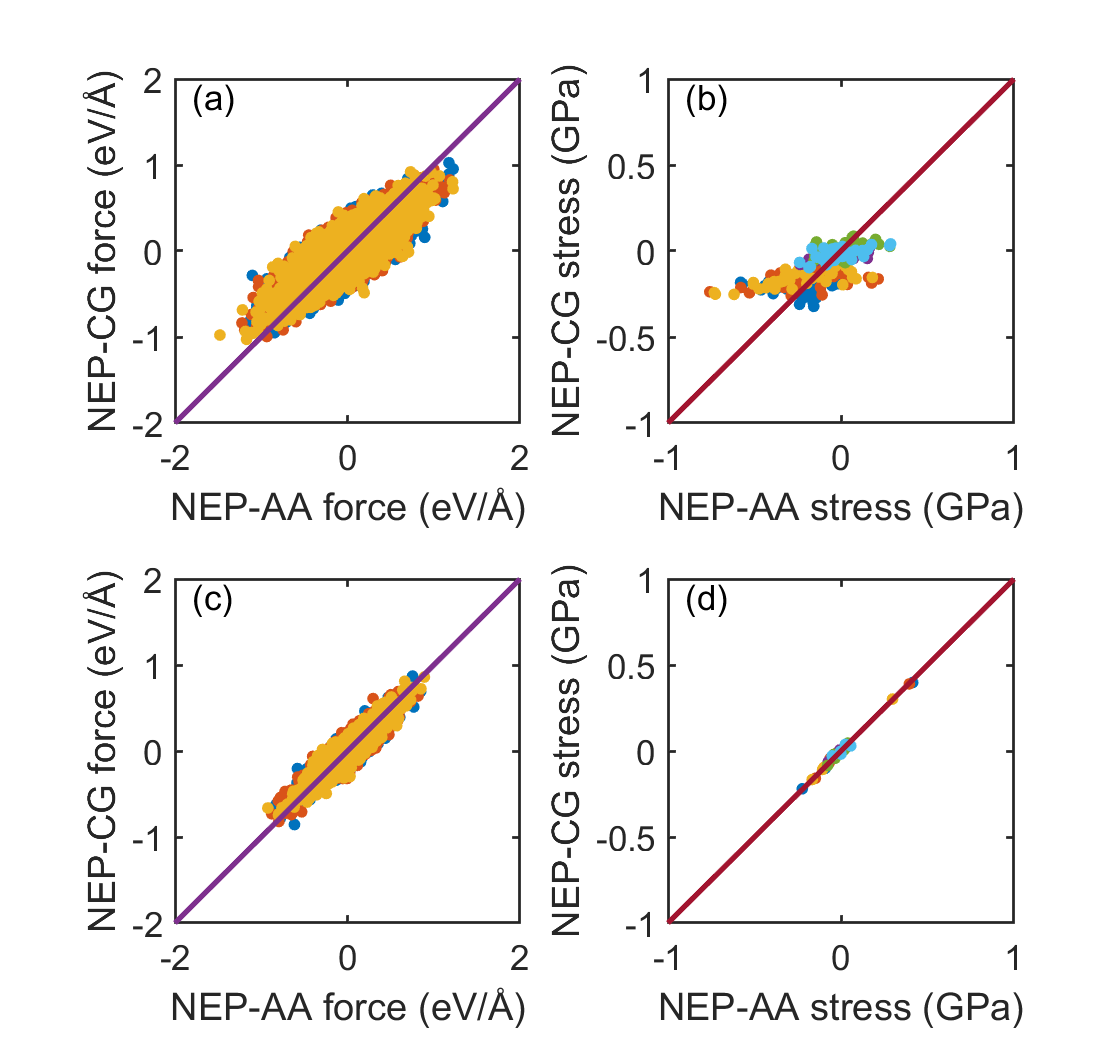}
    \caption{Parity plots comparing forces and stresses predicted by NEP-CG models against NEP-AA reference data for liquid water at 300 K. (a,b) Results from a NEP-CG model trained against instantaneous forces and stresses sampled at 1 bar, showing systematic deviation from the $y=x$ line and significant scatter. (c,d) Results from a NEP-CG model trained against ensemble-averaged forces and stresses using our constrained dynamics approach, incorporating data sampled across target pressures from 1 bar to 0.5 GPa. The ensemble-based model demonstrates excellent agreement and substantially reduced errors across the entire pressure range. Different colors correspond to different force and virial components.}
    \label{fig:parity-water}
\end{figure}

Figure~\ref{fig:parity-water} shows parity plots for forces and stresses. For comparison, we also trained a NEP-CG model using the conventional instantaneous-force method on a NEP-AA trajectory at 300 K and 1 bar. The instantaneous-force approach yields significant deviation from the $y=x$ line with an under-predicted slope (Fig.~\ref{fig:parity-water}a,b), because instantaneous forces are not the correct targets and the true mean forces are buried in noise. This issue is more pronounced for stresses. The resulting \glspl{rmse} are 0.15 eV/\AA~ for forces and 0.14 GPa for stresses.
The force \gls{rmse} is consistent with values reported by Loose \textit{et al.} \cite{loose2023jpcb} for different \glspl{mlp} trained against empirical water models.

In contrast, our ensemble-force training scheme produces parity plots that closely follow the $y=x$ line (Fig.~\ref{fig:parity-water}c,d), confirming that ensemble-averaged forces are the correct targets. The \glspl{rmse} for forces and stresses are reduced to 0.080 eV/\AA~ and 0.0084 GPa, respectively. 
Previous MLP-CG studies typically consider only a single pressure state point, whereas our model accurately predicts stresses across 1 bar to 0.5 GPa, demonstrating transferability across pressure conditions.

\subsubsection{Structural properties and equation of state}

\begin{figure}
    \centering
    \includegraphics[width=\linewidth]{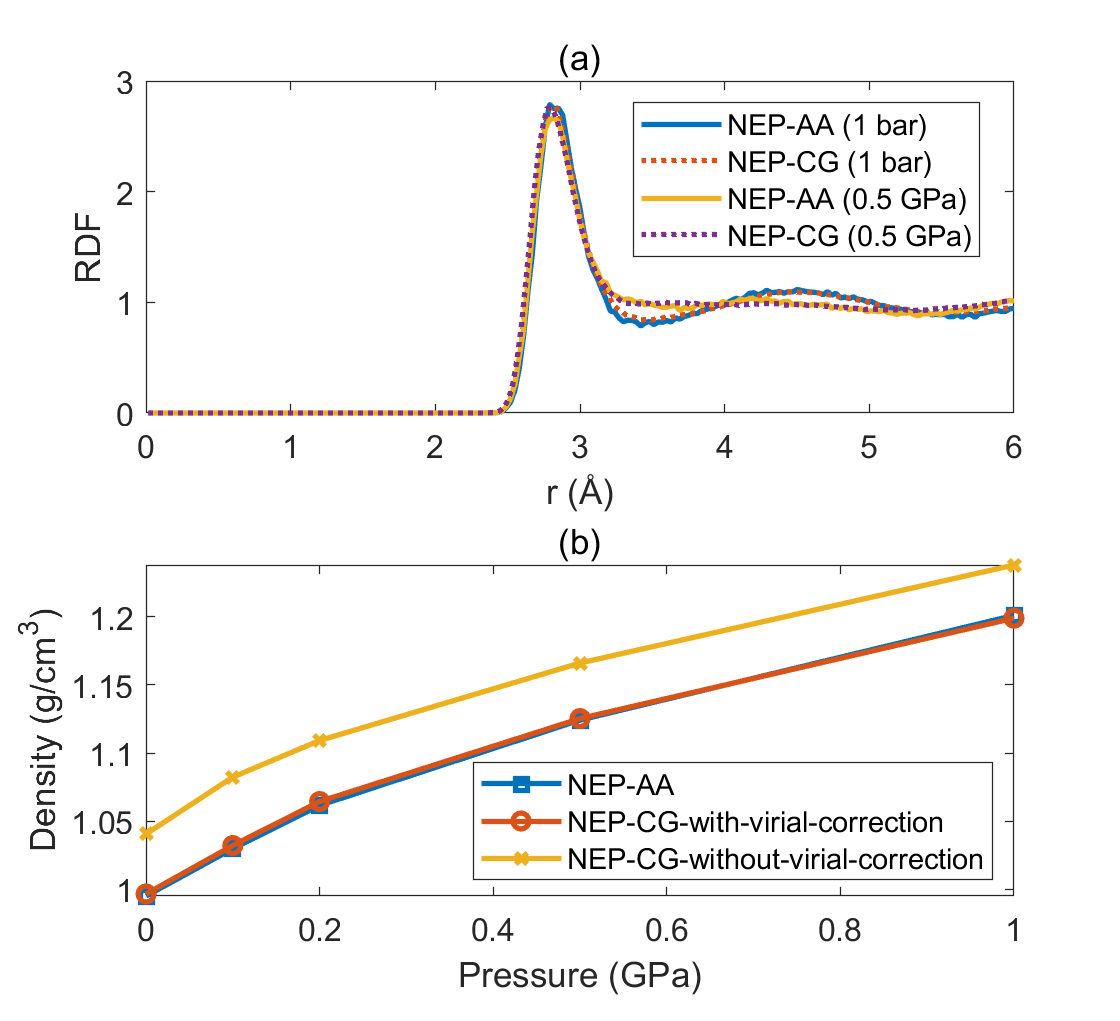}
    \caption{Structural properties and equation of state of liquid water predicted by the NEP-CG model compared to NEP-AA reference data at 300 K. (a) Radial distribution functions $g(r)$ for CG beads at 1 bar and 0.5 GPa, showing excellent structural agreement. (b) Density as a function of pressure from 1 bar to 1 GPa. Results are shown for the NEP-AA reference, the NEP-CG model with virial correction (Eq.~\ref{equation:virial_correction}), and the NEP-CG model without virial correction. The virial-corrected model accurately reproduces NEP-AA densities across the entire pressure range.}
    \label{fig:water-rdf-density}
\end{figure}

A canonical validation of a \gls{cg} model is its ability to reproduce the \gls{rdf} of the beads from the underlying AA model. Figure~\ref{fig:water-rdf-density}a compares \gls{rdf}s calculated using the NEP-CG and NEP-AA models at 1 bar and 0.5 GPa. The excellent agreement demonstrates transferability across different pressure conditions.

Figure~\ref{fig:water-rdf-density}b compares densities from 1 bar to 1 GPa. Training data only extended to 0.5 GPa; the inclusion of 1 GPa tests extrapolation capability. The NEP-CG model shows good agreement with the NEP-AA model across all pressures, with density increasing from approximately 1 g/cm$^3$ at 1 bar to 1.2 g/cm$^3$ at 1 GPa. Successful extrapolation beyond the training range underscores the robustness of our ensemble-based approach.

Figure~\ref{fig:water-rdf-density}b also shows results from a NEP-CG model trained without the virial correction (Eq.~\ref{equation:virial_correction}). Without this correction, density is significantly overestimated across the entire pressure range, confirming that the virial correction is essential for accurately reproducing the equation of state.

\subsection{NEP-CG model for fullerene monolayer network}

Our second example applies the NEP-CG framework to a monolayer of covalently bonded C$_{60}$ molecules in the quasi-hexagonal phase (QHP) \cite{hou2022synthesis}, which exhibits anisotropic mechanical and thermal properties due to directional bonding. An atomistic NEP-AA model for this system was recently developed to study its thermal transport \cite{dong2023ijhmt}. Here we construct a \gls{cg} model where each C$_{60}$ molecule is represented by a single bead.

A key feature of the QHP-C$_{60}$ monolayer is anisotropic bonding: each C$_{60}$ molecule is covalently linked to six neighbors, with bond types differing by direction: \([2+2]\) cycloaddition bonds along the \([010]\) direction ($y$-axis), and C-C single bonds along the \([110]\) and \([1\bar{1}0]\) directions. To capture this anisotropy, we introduce two distinct bead types corresponding to the two C$_{60}$ molecules in the primitive cell. Using a single bead type would not distinguish the $x$ and $y$ directions and would fail to capture the anisotropy. Indeed, correctly encoding species has been demonstrated to be crucial, as neglecting it may introduce unphysical symmetries \cite{gorlich2026jcim}.

\subsubsection{Training data and hyperparameters}

Training data were generated from NEP-AA simulations of a 7200-atom system, mapping onto 120 \gls{cg} beads. For each training structure, we performed NPT simulations at 300 K under target in-plane pressures ranging from $-1$ to $1$ GPa in both directions, using a 1 fs timestep. After 1 ns equilibration, the C$_{60}$ center-of-mass positions were constrained, and ensemble-averaged forces and virials were accumulated over a 10 ns production run. In total, 23 training structures were generated, capturing the anisotropic response under various in-plane strain states.

Due to the large size of each C$_{60}$ molecule and the extended range of intermolecular interactions, a cutoff radius of 25~\AA~ is optimal for both radial and angular descriptors. This relatively large cutoff is necessary to adequately capture interactions between the large C$_{60}$ molecules and ensure accurate representation of the anisotropic bonding environment. Because coarse-graining integrates out intramolecular degrees of freedom, the resulting potential energy surface is substantially smoother, allowing a small neural network with $N_{\mathrm{neu}} = 5$ to achieve high training accuracy.

\subsubsection{Model accuracy and importance of bead-type distinction}

\begin{figure}
    \centering
    \includegraphics[width=\linewidth]{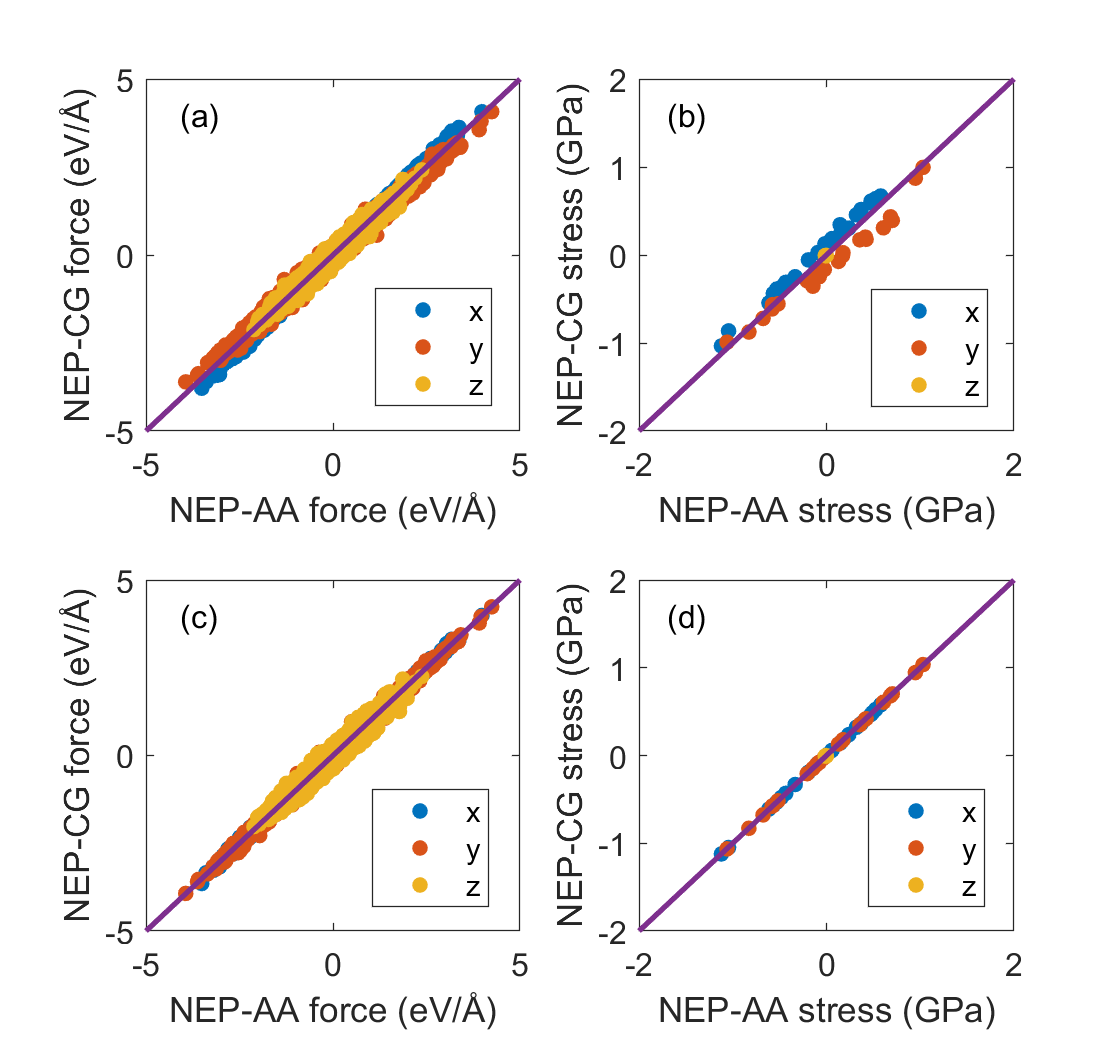}
    \caption{Parity plots comparing forces and stresses predicted by NEP-CG models against NEP-AA reference data for the QHP-C$_{60}$ monolayer at 300 K under various in-plane strain states. (a,b) Results from a one-type NEP-CG model treating all C$_{60}$ molecules as identical beads, exhibiting systematic errors. (c,d) Results from a two-type NEP-CG model distinguishing the two crystallographically distinct C$_{60}$ molecules, accounting for directional dependence of covalent linkages. The two-type model demonstrates substantially improved agreement, particularly for stress components.}
    \label{fig:c60_parity}
\end{figure}

Figure~\ref{fig:c60_parity} presents parity plots comparing forces and stresses predicted by two NEP-CG models. The one-type model, treating all C$_{60}$ molecules as identical beads, exhibits systematic errors (Fig.~\ref{fig:c60_parity}a,b), with \gls{rmse} values of 0.11 eV/\AA~ for forces and 0.083 GPa for stresses. These deviations indicate failure to capture the directional dependence of covalent linkages.

In contrast, the two-type model, distinguishing crystallographically distinct molecules, demonstrates substantially improved agreement (Fig.~\ref{fig:c60_parity}c,d). Force RMSE decreases to 0.090 eV/\AA, while stress RMSE drops by more than an order of magnitude to 0.0025 GPa. These accuracy gains are made observable through our low-noise ensemble-average training scheme.

The superior performance of the two-type model underscores a key principle for coarse-graining anisotropic systems: when atomistic interactions exhibit strong directional dependence that cannot be captured by spherically symmetric beads alone, introducing distinct bead types based on crystallographic equivalence provides a natural way to encode this anisotropy. However, this strategy has limitations, necessitating genuinely anisotropic \gls{cg} models in more general cases \cite{Nguyen2022jcp,wilson2023jcp}.

\subsubsection{Thermal transport}

\begin{figure}
    \centering
    \includegraphics[width=\linewidth]{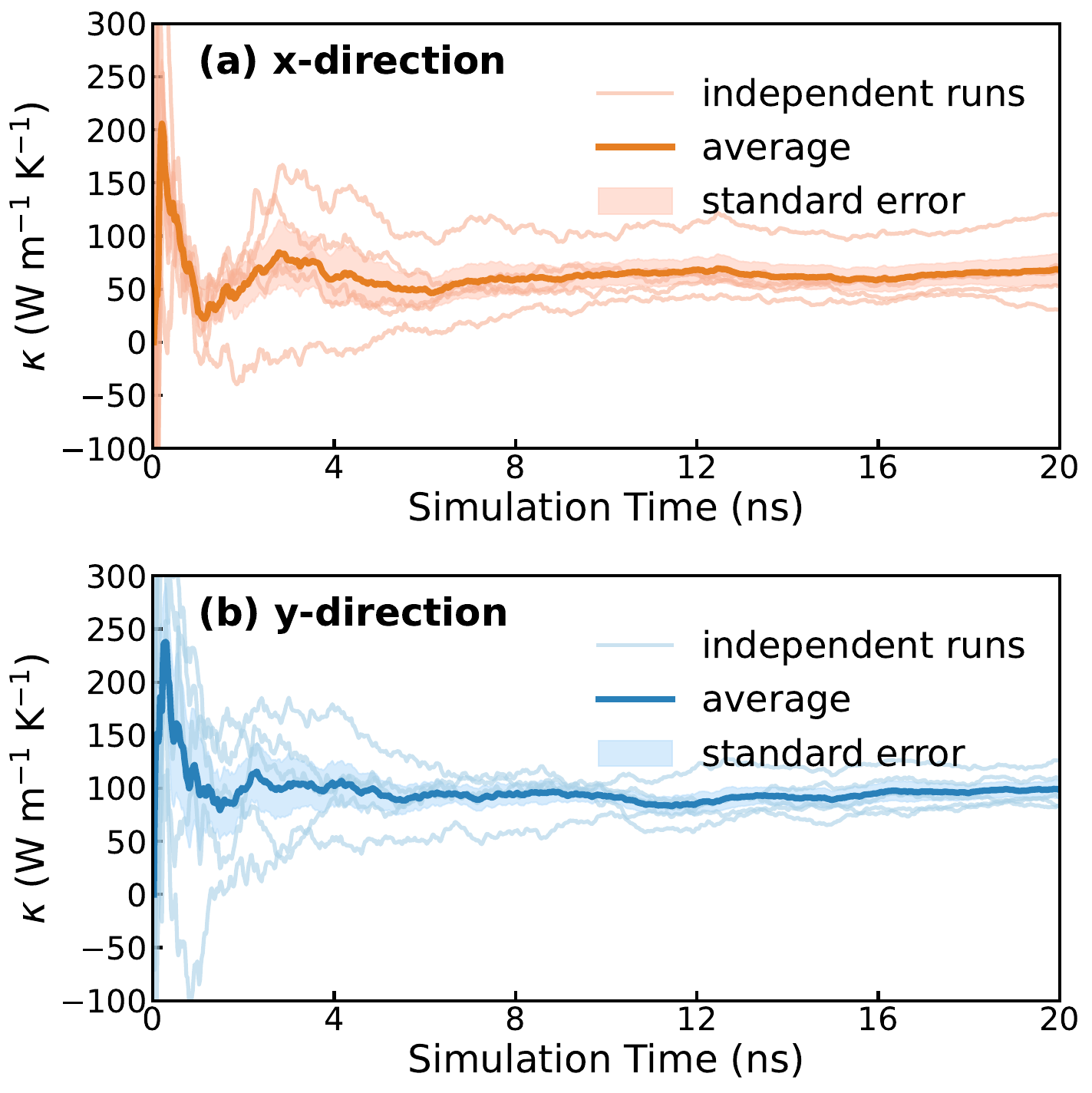}
    \caption{Lattice thermal conductivity of the QHP-C$_{60}$ monolayer as a function of production time in HNEMD simulations. (a) Thermal conductivity along the $x$-direction (average of $[110]$ and $[1\bar{1}0]$ directions). (b) Thermal conductivity along the $y$-direction ($[010]$ direction). Lighter lines show five independent simulations, darker lines their average, and shaded areas the standard error bounds. CG results are scaled by a factor of 60 to account for reduced degrees of freedom.}
    \label{fig:c60_kappa}
\end{figure}

The underlying NEP-AA model was originally developed to study thermal transport in the QHP-C$_{60}$ monolayer \cite{dong2023ijhmt}. We further validate the two-type NEP-CG model by examining its ability to reproduce thermal conductivity using the homogeneous non-equilibrium molecular dynamics (HNEMD) method \cite{fan2019prb}.
The effective thickness of the monolayer is chosen as 8.785 \AA{} to be consistent with the previous NEP-AA calculations \cite{dong2023ijhmt}.

The NEP-CG model successfully reproduces the anisotropic heat conduction characteristic of the QHP-C$_{60}$ monolayer, with thermal conductivity $\kappa$ larger in the $y$-direction than in the $x$-direction. For quantitative comparison, we note that the \gls{cg} model has 60 times fewer degrees of freedom than the atomistic system (one bead per 60 atoms), hence 60 times smaller volumetric heat capacity. Multiplying the raw NEP-CG thermal conductivity values by this factor yields results of the same order as the NEP-AA reference: scaled NEP-CG values are $\kappa_x = 64  \pm 12$ W/mK and $\kappa_y = 95  \pm 9$ W/mK, compared to NEP-AA results of $\kappa_x = 102 \pm 3$ W/mK and $\kappa_y = 137 \pm 7$ W/mK. The CG model thus captures both the qualitative anisotropy and approximate magnitude of thermal conductivity despite the significant reduction in degrees of freedom.

\subsection{NEP-AACG model for gold}

\begin{figure}
    \centering
    \includegraphics[width=\linewidth]{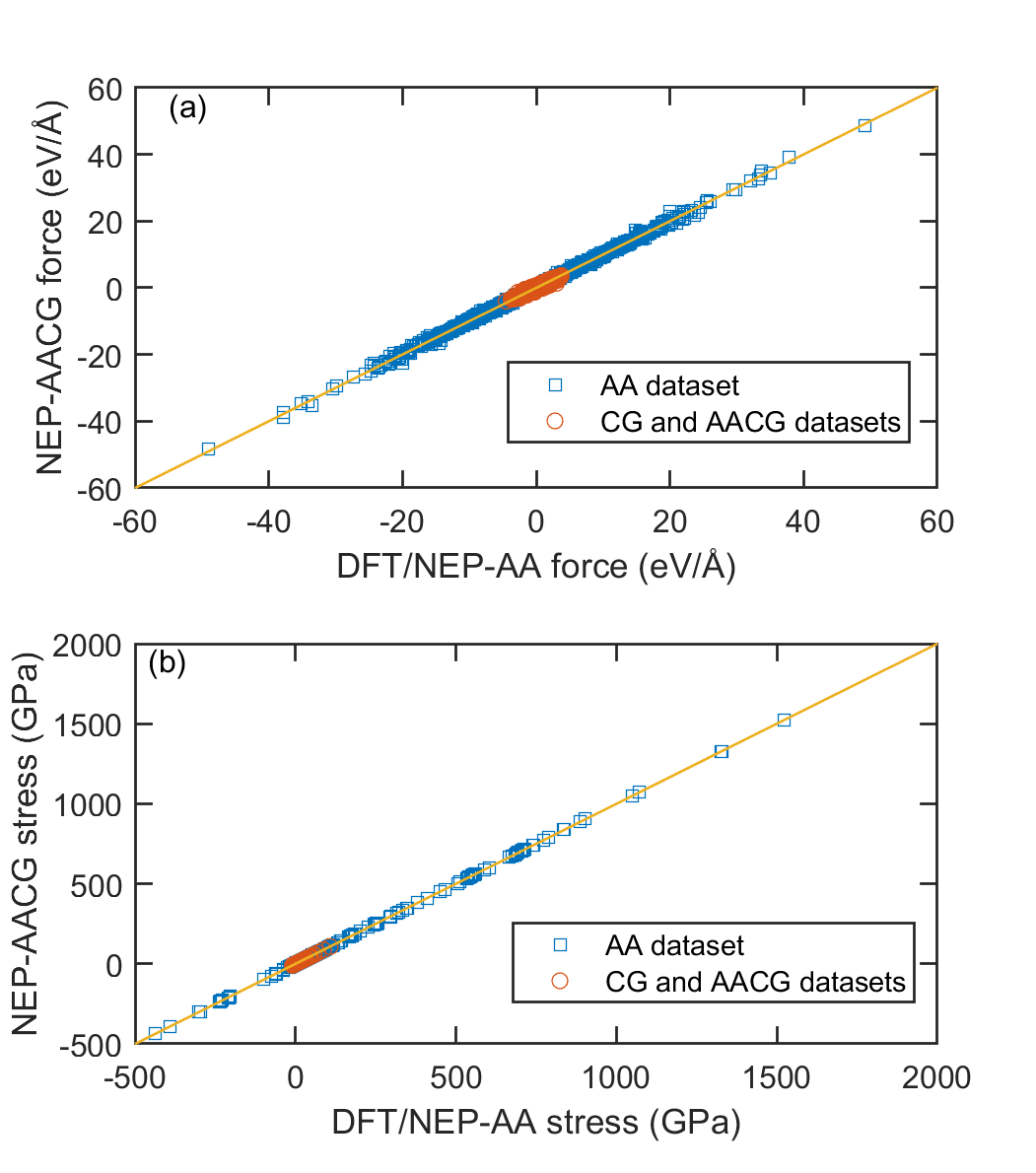}
    \caption{Parity plots for (a) forces and (b) stresses predicted by the unified NEP-AACG model compared to reference data. Squares represent the pure AA dataset (DFT references \cite{song2024nc}). Circles represent both pure CG and mixed AACG datasets (references from constrained NEP-AA simulations). All predictions are from the same NEP-AACG model, demonstrating its ability to simultaneously describe atomistic, coarse-grained, and mixed-resolution configurations.}
    \label{fig:gold_parity}
\end{figure}

Our third case study demonstrates the NEP-AACG framework for gold systems. The underlying NEP-AA model is taken from the UENP-v1 potential \cite{song2024nc}, developed for 16 metals and their alloys, which outperforms traditional embedded-atom method potentials across a wide range of physical properties. We focus on 300 K while considering a comprehensive set of strain conditions.

\subsubsection{Training data generation}

To generate training data, we perform NVT simulations using the NEP-AA model at 300 K under various strain states: isotropic strains \(-10\%\) to \(5\%\), biaxial strains \(-10\%\) to \(5\%\), uniaxial strains \(-10\%\) to \(10\%\), and shear strains \(-5\%\) to \(5\%\). Simulations use 2048 gold atoms with a 5 fs timestep: 1 ns NVT equilibration followed by 10 ns constrained simulation to accumulate reference data.

For each strained configuration, we design two mapping schemes. In pure \gls{cg} mapping, each CG bead represents four gold atoms (one FCC unit cell), reducing 2048 atoms to 512 beads. For mixed-resolution AACG mapping, half the atoms are coarse-grained using the same mapping, while the remaining atoms retain atomistic resolution, enabling multiscale simulations.

\subsubsection{Model hyperparameters}

We train a NEP-AACG model by combining three datasets: the original Au dataset from UENP-v1, the pure CG dataset, and the AACG dataset. 
Within the NEP framework, we define two particle types: atomistic Au and CG beads, which have different effective sizes.
Based on systematic testing, using different cutoff radii for different particle types proves beneficial. For the original UENP-v1 model, radial and angular cutoffs are 6~\AA~ and 5~\AA, respectively. For CG beads, larger cutoffs of 8~\AA~ and 7~\AA~ are optimal, reflecting their larger effective size. For mixed atom-bead interactions, we define cutoffs as arithmetic means of pure-type cutoffs: 7~\AA~ (radial) and 6~\AA~ (angular). With training data encompassing both pure AA and mixed-resolution configurations, we select an intermediate hidden-layer size $N_{\rm neu}=30$, balancing model capacity and computational efficiency. 

\subsubsection{Model accuracy}

Figure~\ref{fig:gold_parity} presents parity plots for forces and stresses predicted by the unified NEP-AACG model. The model demonstrates excellent accuracy across all three datasets. For the pure AA dataset (\gls{dft} references \cite{song2024nc}), the NEP-AACG model achieves \glspl{rmse} of 0.10 eV/\AA~ for forces and 0.50 GPa for stresses, comparable to the original UENP-v1 model \cite{song2024nc}, confirming that adding CG and mixed-resolution data does not compromise atomistic accuracy.

For pure CG and mixed AACG configurations (references from constrained NEP-AA simulations), force RMSE is 0.077 eV/\AA~ and stress RMSE is 0.086 GPa. The substantially lower stress error reflects both the smoother CG potential energy surface and the narrower range of forces and stresses in CG-related datasets.

Crucially, all predictions are generated by a single unified NEP-AACG model with two particle types and interaction-specific cutoffs. Simultaneous accuracy across purely atomistic, purely coarse-grained, and mixed-resolution configurations demonstrates that the model has learned a consistent free energy surface spanning multiple resolution levels, enabling seamless multiscale simulations without interface artifacts.

\subsubsection{Bulk mechanical validation}

\begin{figure}
    \centering
    \includegraphics[width=\linewidth]{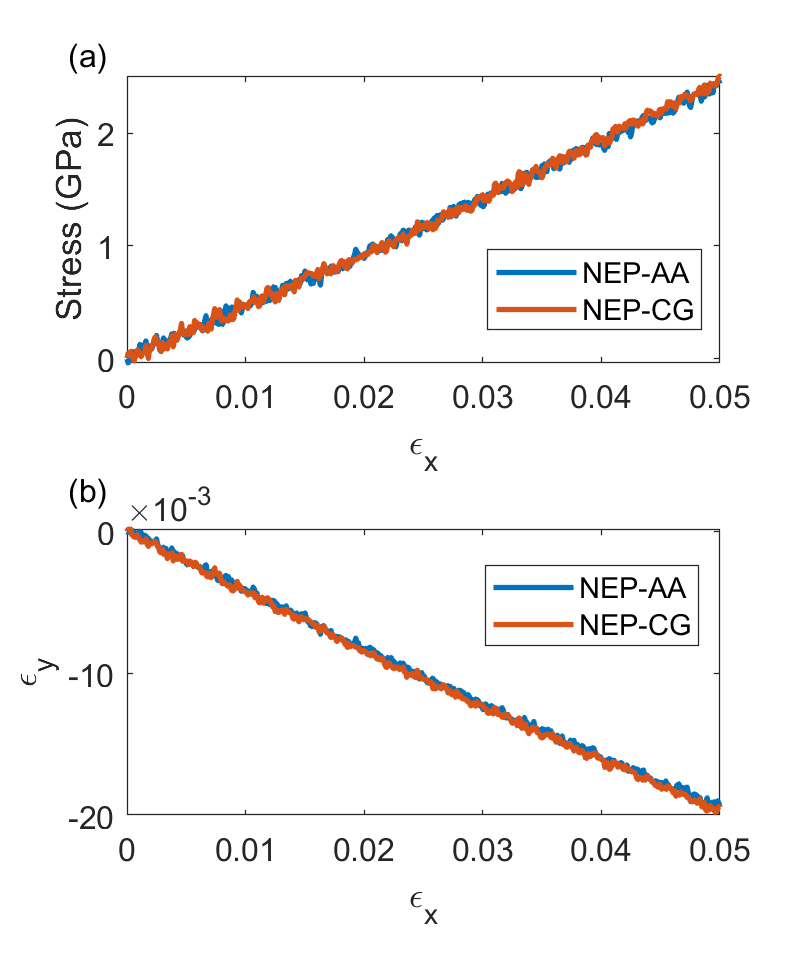}
    \caption{Validation of the NEP-AACG model for uniaxial tensile deformation of bulk gold. (a) Axial stress $\sigma_x$ versus applied engineering strain $\varepsilon_x$ for fully atomistic NEP-AA and pure CG NEP-AACG models. (b) Transverse strain $\varepsilon_y$ versus axial strain $\varepsilon_x$, illustrating the Poisson effect. Excellent agreement across the strain range up to $5\%$ confirms that the coarse-grained model faithfully reproduces the mechanical response. Simulations at 300 K.}
    \label{fig:stress-strain}
\end{figure}

To validate predictive capability, we compare the NEP-AACG model against the reference NEP-AA model in uniaxial tensile deformation of bulk face-centered cubic gold along the $x$-direction, with engineering strain $\varepsilon_x$ up to 5\%. Two simulations are conducted: fully atomistic simulation with NEP-AA and 16384 atoms, and pure CG simulation with NEP-AACG and 4096 beads.

Figure~\ref{fig:stress-strain}a shows the stress-strain behavior. Axial stress $\sigma_x$ increases linearly with $\varepsilon_x$ in the elastic regime. Transverse strain $\varepsilon_y$ exhibits the expected Poisson contraction (Fig.~\ref{fig:stress-strain}b). From the linear region, we estimate a Poisson ratio of approximately $\nu \approx 0.4$, consistent with typical values for gold. The NEP-AACG results agree quantitatively with the NEP-AA reference across the entire strain range, confirming that the coarse-grained model, trained on ensemble-averaged data from strained configurations, successfully captures the mechanical response.

\subsubsection{Multiscale nanowire fracture simulation}

\begin{figure*}
    \centering
    \includegraphics[width=0.8\linewidth]{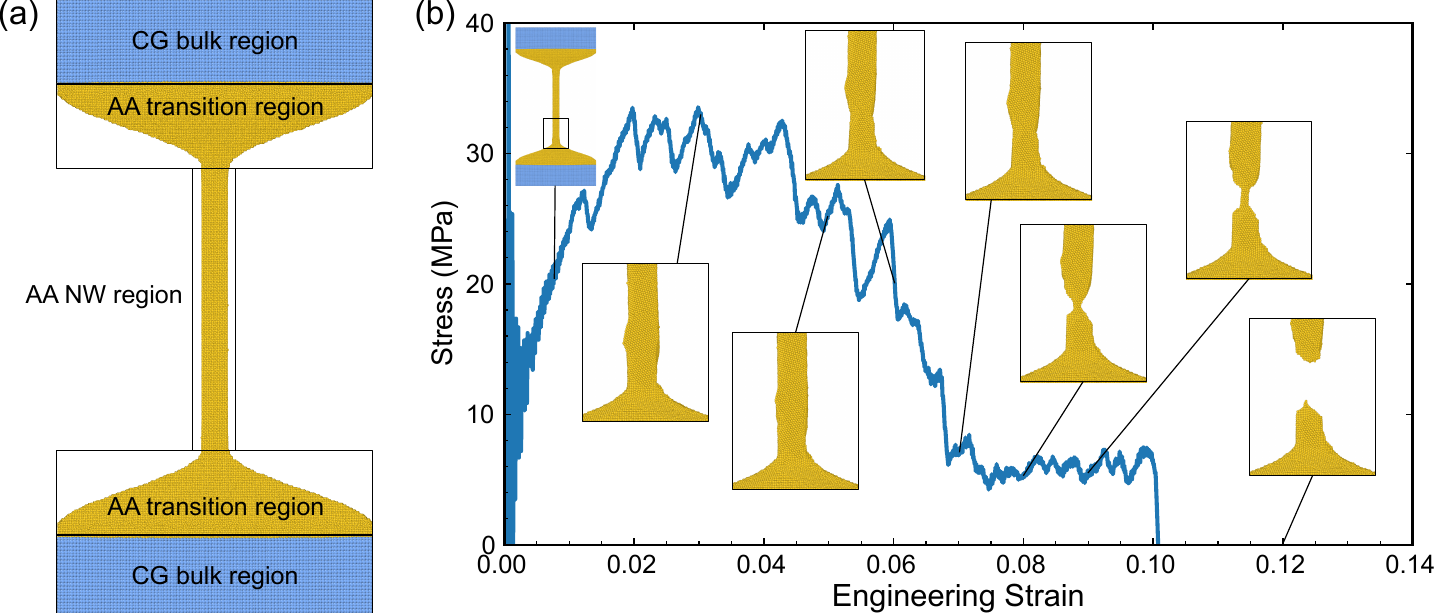}
    \caption{Tensile deformation of a gold nanowire using the NEP-AACG model. (a) Schematic of the tensile-test setup: CG bulk reservoir, atomistic AA transition zone, and atomistic nanowire specimen. (b) Engineering stress–strain response. Insets show representative atomic configurations at selected strains for the area highlighted by the black rectangle.}
    \label{fig:nanowire}
\end{figure*}

After validating the NEP-AACG model for bulk simulations, we demonstrate its capabilities in a proof-of-concept multiscale application: simulating fracture of gold nanowires under tension, which has attracted extensive interest \cite{jorgensen2025acsnano}. This problem exemplifies the multiscale nature of materials deformation, where atomistic resolution is required in the fracture region while extended length scales are necessary to avoid boundary condition artifacts.

We construct a gold nanowire tensile-test model with a total length of approximately 80 nm, comprising a few regions: a central atomistic nanowire region of about 37 nm long and 3 nm thick where fracture is expected, two end regions modeled as CG beads, and two AA transition regions. The CG regions extend to the boundaries, serving as reservoirs that apply realistic mechanical constraints to the atomistic zone (Fig.~\ref{fig:nanowire}a). 

We apply uniaxial tension by deforming the system at a constant engineering strain rate of \(10^7\) s\(^{-1}\), which is within the range achievable in high-speed loading experiments \cite{sun2022material}. The simulated stress-strain relation is shown in Fig.~\ref{fig:nanowire}b, along with snapshots at different strain levels.

The maximum stresses fluctuate about 30 MPa at strains between 0.01 and 0.05, during which the nanowire maintains its overall structural integrity. Beyond a strain of 0.07, the stress drops to below 10 MPa as the nanowire begins to thin in the middle. Finally, at strains exceeding 0.1, the nanowire undergoes complete fracture.

\subsection{Computational performance gains}

A key advantage of coarse-grained models is computational efficiency, enabling access to extended time and length scales. We benchmark NEP-CG models against their NEP-AA counterparts using two metrics: particle throughput (particle\(\cdot\)step/s) and effective simulation speed (ns/day) for equivalent spatial sizes. Benchmarks were performed on an NVIDIA RTX 5090 GPU using GPUMD \cite{xu2025mgea}, with each simulation running 1000 steps in the NVT ensemble across various system sizes.

Particle throughput isolates intrinsic computational complexity, primarily influenced by neural network size and average number of neighbors per particle. For water, the NEP-CG model achieves significantly higher throughput due to its smaller network ($N_{\mathrm{neu}}=10$ vs. $60$) and fewer neighbors per bead from reduced particle density. For C$_{60}$, the advantage is even more pronounced: the NEP-CG model uses $N_{\mathrm{neu}}=5$ versus $N_{\mathrm{neu}}=50$ used for the NEP-AA model, and the 25~\AA{} cutoff, while large in absolute value, still yields fewer neighbors per bead due to reduced interaction sites.

More practically relevant is effective simulation speed in ns/day for the same spatial size, incorporating three factors favoring CG models: higher particle throughput, larger allowable timesteps due to smoother CG energy surfaces and elimination of high-frequency vibrations, and fewer beads than atoms for the same volume.

For water, the NEP-AA model uses a 0.5 fs timestep; the NEP-CG model safely accommodates 2 fs. Combined with higher throughput and 3:1 atom-to-bead reduction, the NEP-CG model achieves approximately 50-fold speedup in ns/day for equivalent spatial sizes.

For the C$_{60}$ monolayer, gains are even more dramatic. The NEP-AA model requires a 1 fs timestep; the NEP-CG model allows 20 fs. This, combined with 60:1 atom-to-bead reduction and substantially higher throughput, yields approximately 1000-fold speedup in ns/day for equivalent spatial sizes.

\begin{figure}
    \centering
    \includegraphics[width=\linewidth]{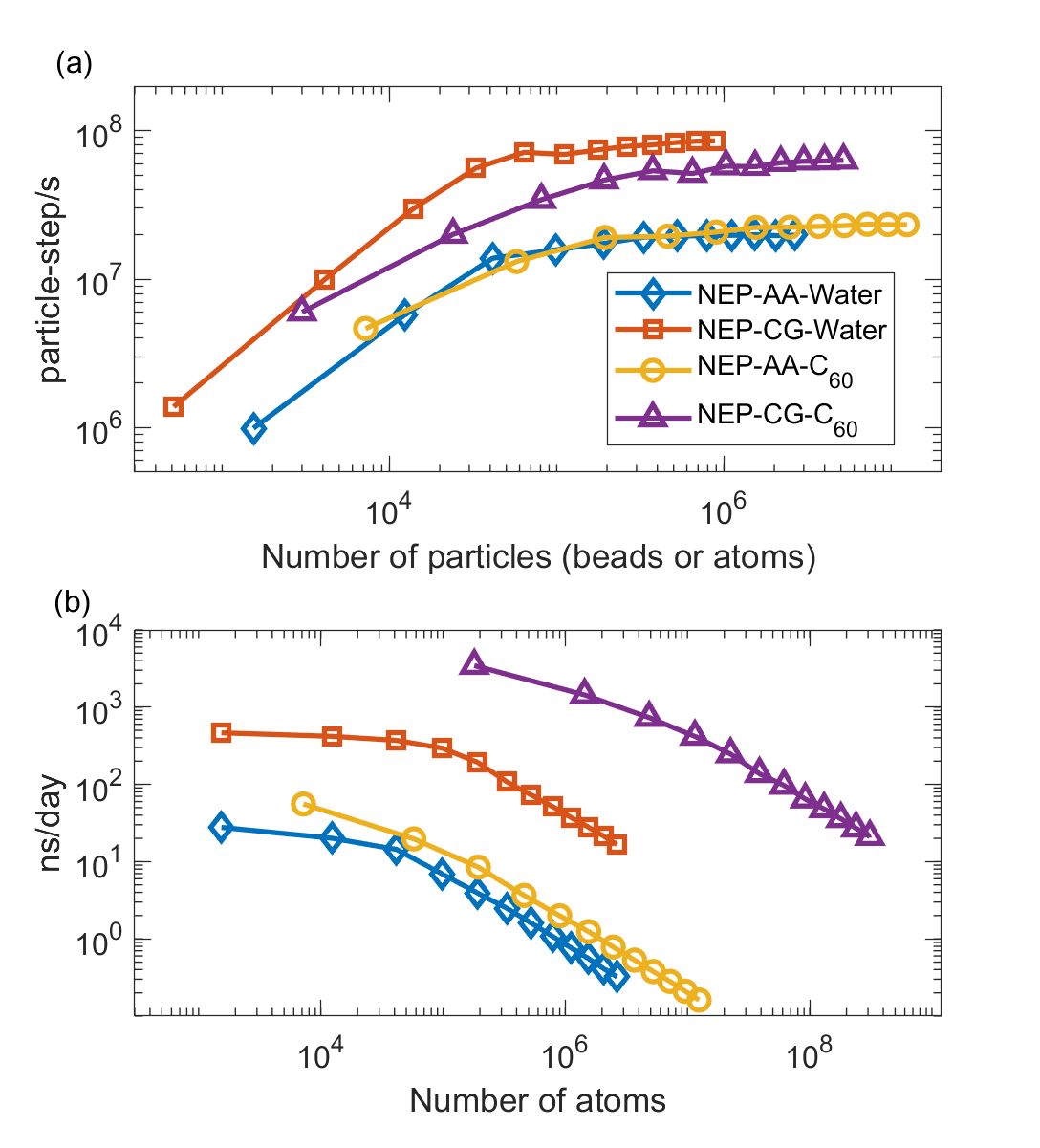}
    \caption{Computational performance comparison between NEP-CG and NEP-AA models. (a) Particle throughput (particle\(\cdot\)step/s) versus number of particles, demonstrating higher intrinsic efficiency of CG models due to smaller neural networks and fewer neighbors per particle. (b) Effective simulation speed (ns/day) versus equivalent atom count, accounting for larger timesteps and reduced particle count. Benchmarks on NVIDIA RTX 5090 GPU with GPUMD.}
    \label{fig:speed}
\end{figure}

\section{Summary and Conclusions}

We have developed and demonstrated two complementary methods for constructing coarse-grained models within the \gls{nep} framework: NEP-CG for pure coarse-grained simulations and NEP-AACG for multiscale simulations that seamlessly integrate atomistic and coarse-grained degrees of freedom. The central innovation of our approach lies in generating low-noise training data through constrained molecular dynamics simulations, which directly yield ensemble-averaged forces corresponding to the potential of mean force. This contrasts with conventional force-matching methods that fit to noisy instantaneous forces, often leading to large training errors and low data efficiency.

We illustrated the capabilities of our methods through three representative examples. For liquid water, the NEP-CG model trained with ensemble-averaged forces achieved substantially lower errors than models trained on instantaneous forces, with force and stress RMSE reductions of nearly 50\% and an order of magnitude, respectively. A virial correction scheme compensating for lost degrees of freedom proved essential for accurately reproducing the density-pressure equation of state across pressures from 1 bar to 1 GPa, including successful extrapolation beyond the training range.

For the quasi-hexagonal phase C$_{60}$ monolayer, we demonstrated the importance of capturing anisotropic bonding through distinct bead types corresponding to crystallographically inequivalent molecules. The two-type NEP-CG model dramatically improved stress prediction (RMSE reduction from 0.083 GPa to 0.0025 GPa) compared to a one-type model and successfully reproduced the anisotropic thermal conductivity. After appropriate scaling for reduced degrees of freedom, the CG model yielded thermal conductivities of the same order as the atomistic reference.

Our third example showcased the NEP-AACG framework for gold, demonstrating its ability to simultaneously describe purely atomistic, purely coarse-grained, and mixed-resolution configurations within a single unified model. The model accurately reproduced the stress-strain behavior of bulk gold under uniaxial tension and enabled multiscale simulations of nanowire fracture at an experimentally relevant strain rate of \(10^7\) s\(^{-1}\).

Finally, we quantified the computational gains enabled by our NEP-CG models. For water, the combination of higher particle throughput, a fourfold timestep increase (0.5 fs to 2 fs), and a 3:1 particle reduction yielded approximately 50\(\times\) speedup in ns/day. For the C$_{60}$ monolayer, a 20-fold timestep increase (1 fs to 20 fs) combined with a 60:1 particle reduction and higher throughput resulted in a dramatic 1000\(\times\) speedup.
Overall, the computational speeds for NEP-CG models reach hundreds to thousands of ns/day.

Despite these successes, our work has several limitations that point to important future directions. First, all training data were generated at a single temperature (300 K). The potential of mean force is inherently temperature-dependent, and extending our approach to incorporate multiple temperatures \cite{ruza2020jcp} represents a natural step toward thermodynamically consistent CG models.

Second, our models employ only isotropic beads. While the two-type approach for the C$_{60}$ monolayer successfully captured mechanical and thermal anisotropy, this strategy cannot represent situations where orientational degrees of freedom are essential \cite{Nguyen2022jcp,wilson2023jcp,campos2024npjcm,argun2025jcp}. Extending the NEP framework to accommodate anisotropic particles would significantly broaden applicability to soft matter systems such as polymers and liquid crystals.

Third, the partitioning into \gls{aa} and \gls{cg} regions in our NEP-AACG models is predefined and fixed during simulations, without the capability for dynamic resolution adaptation. This limits applications where the required level of detail changes over time, such as in systems where important events may occur in initially coarse-grained regions.

Future work will explore these directions, aiming to develop temperature-transferable CG models and anisotropic bead representations within the NEP-CG and NEP-AACG frameworks. Applications to more complex systems, including biomolecules and soft materials, will further demonstrate the versatility and power of our ensemble-based training methodology.

\vspace{0.5cm}
\noindent \textbf{Data availability:}
All the training datasets and trained machine-learned potential models generated in this work are freely available in the nep-data repository (\url{https://gitlab.com/brucefan1983/nep-data}).

\vspace{0.5cm}
\noindent \textbf{Code availability:}
All the calculations in this work can be performed with  GPUMD-v4.9.1 and future versions (\url{https://github.com/brucefan1983/GPUMD}).

\begin{acknowledgments}
This work was supported by the Advanced Materials-National Science and Technology Major Project (No. 2024ZD0606900). 
ZF was supported by the Science Foundation from Education Department of Liaoning Province (No. LJ232510167001).
KX acknowledges support from the Department of Science and Technology of Liaoning Province (No. JYTMS20231613) and Bohai University On-campus Doctoral Start-up Project Funding.

\end{acknowledgments}

\section*{Declaration of Conflict of Interest}
The authors have no conflicts to disclose.

\end{document}